\documentclass[12pt]{article}
\usepackage[latin1]{inputenc}
\usepackage[dvips]{graphicx}
\newcommand{\exprm}{{\rm exp}}

\newcommand{\pa}{\partial}

\newcommand{\text}{\rm}

\newcommand{\ug}{ \; = \; }

\newcommand{\infi}{\infty}

\newcommand{\la}{\lambda}
\newcommand{\kr}{k_{\rho}}

\newcommand{\bb}{\begin{equation}}
\newcommand{\ee}{\end{equation}}
\newcommand{\bega}{\begin{eqnarray}}
\newcommand{\ega}{\end{eqnarray}}
\newcommand{\begae}{\begin{eqnarray*}}
\newcommand{\egae}{\end{eqnarray*}}

\newcommand{\h}{\hspace*{4ex}}
\newcommand{\dis}{\displaystyle}

\newcommand{\be}{\beta}

\newcommand{\om}{\omega}

\newcommand{\cent}{\centerline}
\newcommand{\vs}{\vspace*}

\begin{document}

\baselineskip 0.8cm

\begin{center}

{\large {\bf Chirped optical X-shaped pulses in material media}$^{\:
(\dag)}$} \footnotetext{$^{\: (\dag)}$  Work partially supported
by FAPESP (Brazil), and by MIUR-MURST and INFN (Italy); previously
available as e-print physics/0405059. \ E-mail addresses for
contacts: mzamboni@dmo.fee.unicamp.br}

\end{center}

\vs{3mm}

\cent{ Michel Zamboni-Rached, }

\vs{0.1 cm}

\centerline{{\em Department of Microwaves and Optics, School of
Electrical Engineering (FEEC),}} \centerline{{\em State University
of Campinas, Campinas, SP, Brasil.}}

\vs{0.4 cm}

\cent{Hugo E. Hern\'andez-Figueroa}

\vs{0.4 cm}

\centerline{{\em Department of Microwaves and Optics, School of
Electrical Engineering (FEEC),}} \centerline{{\em State University
of Campinas, Campinas, SP, Brasil.}}
%%D.M.O., Faculty of Electrical Engineering, UNICAMP, Campinas, SP,
%%Brasil.}}

\vs{0.4 cm}

\cent{ Erasmo Recami }

\vs{0.1 cm}

\cent{{\em Facolt\`a di Ingegneria, Universit\`a statale di
Bergamo, Dalmine (BG), Italy;}} \cent{{\em {\rm and} \
INFN---Sezione di Milano, Milan, Italy.}}

\vs{0.5 cm}

\

{\bf Abstract  \ --} \ In this paper we analyze the properties of chirped
optical X-shaped pulses
propagating in material media without boundaries. We show that such
("superluminal") pulses may recover their transverse and longitudinal shape
after some propagation distance, while the ordinary chirped gaussian-pulses
can recover their longitudinal shape only (since gaussian pulses suffer a
progressive spreading during their propagation). We therefore propose the use
of chirped optical X-type pulses to overcome the problems of both dispersion
and diffraction during the pulse propagation.\\

%%{\em PACS nos.}: \ 41.20.Jb ; \ 03.50.De ; \ 03.30.+p ; \
%%84.40.Az ; \ 42.82.Et ; \ 83.50.Vr ; \ \ 62.30.+d ; \ 43.60.+d ;
%%\  91.30.Fn ; \  04.30.Nk ; \  42.25.Bs ; \ 46.40.Cd ; \ 52.35.Lv
%%\ .\hfill\break
%%% ??????


\newpage

{\bf 1. -- Introduction}\\

\h Today, the theoretical and experimental existence of {\em localized}
(nondiffracting) solutions to the wave equation in free space
is a well established fact. The corresponding waves propagate for
long distances resisting the diffraction effects, i.e., they
maintain their shape during propagation; and the (Superluminal)
X-shaped waves are examples of these solutions.

\h The theory of the localized waves (LWs) was initially developed
for free space (vacuum)[1-4] and in some situations for waveguides
(hollow or coaxial cables)[5]. Subsequently, the theory was
extended in order to have undistorted wave propagation in material
media without boundaries[6-13]. In this case the LWs are capable
of overcoming both the difraction and dispersion problems for long
distances. The extension of the LW theory to material media was
obtained by making the {\em axicon angle} of the Bessel beams
(BBs) vary with the frequency[6-13], in such a way that a
(frequency) superposition of such BBs does compensate for the
material dispersion.

\h In spite of such an idea to work well in theory[8] and in its
possible experimental implementation[6,7], it is not a simply
realizable one, and requires having recourse to holographic
elements.

\h In this paper we propose a simpler way to obtain pulses capable
of recovering their spatial shape, both transversally and
longitudinally, after some propagation. It consists in
using {\em chirped} optical X-type pulses, while keeping the axicon angle
fixed. Let us recall that, by contrast, chirped gaussian pulses in
unbounded material media may recover only their longitudinal
width, since they undergo a progressive transverse spreading while
propagating.

\newpage

{\bf 2. -- Chirped optical X-shaped pulses in material media}\\

\h Let us start with an axis-symmetric Bessel beam in a material
medium with refractive index $n(\om)$:

\

\bb \psi(\rho,z,t)\ug J_0(k_{\rho}\rho)e^{i\be z}e^{-i\om t} \; ,
\label{bb}\ee

\

where it must be obeyed the condition
$k_{\rho}^2=n^2(\om)\om^2/c^2 - \be^2$, which connects amongs themselves
the transverse and longitudinal wave numbers $k_{\rho}$ and $\be$, and
the angular frequency $\om$.  In addition, we impose that
$k_{\rho}^2 \geq 0$ and $\om / \be \geq 0$, to avoid a nonphysical
behavior of the Bessel function $J_0(.)$ and to confine ourselves
to forward propagation only.

\h Once the conditions above are satisfied, we have the liberty of
writing the longitudinal wave number as $\be = (n(\om)\om
\cos\theta)/c$ and, therefore, $k_{\rho} = (n(\om)\om
\sin\theta)/c$; where (as in the free space case) $\theta$ is the
axicon angle of the Bessel beam.

\h Now we can obtain a X-shaped pulse by performing a frequency
superposition of these BBs, with $\be$ and $k_{\rho}$ given by the
previous relations:

\

\bb \Psi(\rho,z,t) \ug \int_{-\infi}^{\infi}\,
S(\om)\,J_0\left(\frac{n(\om)\om}{c}
\sin\theta\,\rho\right)\,e^{i\be(\om)z}e^{-i\om t}\,d\om \; ,
\label{geral}\ee

\

where $S(\om)$ is the frequency spectrum, and the axicon angle is
kept constant.

\h One can see that the phase velocity of each BB in our
superposition (\ref{geral}) is different, and given by $V_{\rm
phase} = c/(n(\om)\cos\theta)$. So, the pulse given by
Eq.(\ref{geral}) will suffer a dispersion during its propagation.

\h The method developed by S\~onajalg et al.[6,7] and explored by
others[8-13], to overcome this problem, consisted in regarding the
axicon angle $\theta$ as a function of the frequency, in order to
obtain a linear relationship between $\be$ and $\om$.

\h Here, however, we want to work with a {\em fixed} axicon angle, and
we have to find out another way for avoiding dispersion and
diffraction along a certain propagation distance. To do that, we might
choose a chirped Gaussian spectrum  $S(\om)$ in Eq.(\ref{geral}):

\

\bb S(\om) \ug
\frac{T_0}{\sqrt{2\pi(1+iC)}}\,\,e^{\dis{-q^2(\om-\om_0)^2}}
\label{S}\;\;\;\;{\rm with}\;\;\;\;\; q^2 \ug
\frac{T_0^2}{2(1+iC)} \; , \label{s}\ee

\

where $\om_0$ is the central frequency of the spectrum, $T_0$ is a
constant related with the initial temporal width,  and $C$ is the
chirp parameter.  Unfortunately, there is no analytical solution to
Eq.(2) with $S(\om)$ given by Eq.(3), so that some approximations
are to be made.

\h Then, let us assume  that the spectrum $S(\om)$, in the surrounding
of the carrier frequency $\om_0$ , is enough narrow that
$\Delta\om/\om_0<<1$, so to ensure that $\be(\om)$ can be
approximated by the first three terms of its Taylor expansion in
the vicinity of $\om_0$: That is,  $\be(\om)\approx \be(\om_0) +
\be'(\om)|_{\om_0}\,(\om - \om_0) + (1/2) \be''(\om)|_{\om_ 0}\,
(\om - \om_0)^2$; where, after using $\be =n(\om)\om \cos\theta/c$,
it results that
\

\bb \frac{\pa \be}{\pa\om} \ug \frac{\cos\theta}{c}\left[n(\om) +
\om\,\frac{\pa n}{\pa\om} \right]\;\; ; \;\; \frac{\pa^2
\be}{\pa\om^2} \ug \frac{\cos\theta}{c}\left[ 2\frac{\pa
n}{\pa\om} + \om \frac{\pa^2 n}{\pa\om^2} \right] \; . \label{b1}\ee

\

\h As we know, $\be'(\om)$  is related to the pulse group-velocity
by the relation $Vg = 1/ \be'(\om)$.  Here we can see the
difference between the group-velocity of the X-type pulse (with a
fixed axicon angle) and that of a standard gaussian pulse.  Such a
difference is due to the factor $\cos\theta$ in Eq.(4).  Because
of it, the group-velocity of our X-type pulse is always greater
than the gaussian one. In other words, $(V_g)_X =
(1/\cos\theta)(V_g)_{gauss}$.

\h We also know that the second derivative of $\be(\om)$ is
related to the group-velocity dispersion (GVD) $\be_2$ by $\be_2 =
\be''(\om)$.

\h The GVD is responsible for the temporal (longitudinal)
spreading of the pulse. Here one can see that the GVD of the
X-type pulse is always smaller than that of the standard Gaussian
pulses, due the factor $\cos\theta$ in Eq.(\ref{b1}).  Namely:
$(\be_{2})_X = \cos\theta (\be_{2})_{\rm gauss}$.

\h On using the above results, we can write

\

\bb \begin{array}{clcr}
 \Psi(\rho,z,t) \!\!&= \dis{\frac{T_0\,\,e^{i\be(\om_0)z}e^{-i\om_0
t}}{\sqrt{2\pi(1+iC)}}
\,\int_{-\infi}^{\infi}\,J_0\left(\frac{n(\om)\om}{c}
sin\theta\,\rho\right)\times}\\
\\
&\;\;\;\times\,\dis{\exprm\left\{i\frac{(\om-\om_0)}{V_g}\left[z -
V_g t \right]
\right\}\exprm\left\{(\om-\om_0)^2\left[\frac{i\be_2}{2}z - q^2
\right] \right\}\,d\om} \; . \label{geral2} \end{array} \ee

\

\h The integral in Eq.(\ref{geral2}) cannot be solved
analytically, but it is enough for us to obtain the pulse
behavior. Let us analyze the pulse for $\rho=0$. In this case we
obtain:

\

\bb
\Psi(\rho=0,z,t) \ug
\dis{e^{i\be(\om_0)z}e^{-i\om_0 t}\,\frac{T_0}{\sqrt{T_0^2 -
i\be_2(1+iC)z}}\,\exprm\left[\frac{-(z-V_gt)^2(1+iC)}{2V_g^2[T_0^2
- i\be_2(1+iC)z]}\right]} \; . \label{sol1} \ee

\

\h From Eq.(\ref{sol1}) we can immediately see that the initial
temporal width of the pulse intensity is $T_0$ and that, after
some propagation distance $z$, the time-width $T_1$ becomes

\

\bb \frac{T_1}{T_0} \ug \left[\left(1+\frac{C\be_2z}{T_0^2}
\right)^2 +
\left(\frac{\be_2z}{T_0^2}\right)^2\right]^{1/2} \; . \label{T1}\ee

\

\h Relation (\ref{T1}) describes the pulse spreading-behavior. One
can easily show that such a  behavior depends on the sign
(positive or negative) of the product $\be_2C$, as is well known
for the standard gaussian pulses[14].

\h In the case $\be_2C > 0$,  the pulse will monotonically become
broader and broader with the distance $z$.  On the other hand, if
$\be_2C < 0$ the pulse will suffer, in a first stage, a narrowing,
and then it will spread during the rest of its propagation. So,
there will be a certain propagation distance in correspondence to
which the pulse will recover
its initial temporal width ($T_1=T_0$). From relation (\ref{T1}),
we can find this distance $Z_{T1=T_0}$ (considering $\be_2 C <
0$) to be

\

\bb Z_{T_1=T_0} \ug \frac{-2CT_0^2}{\be_2(C^2+1)} \; . \label{Z} \ee

\

\h One may notice that the maximum distance at which our chirped
pulse, with given $T_0$ and $\be_2$,  may recover its initial
temporal width can   be easily evaluated from Eq.(\ref{Z}), and is
results to be $L_{\rm disp} = T_0^2 / \be_2$.  We'll call such a maximum
value $L_{\rm disp}$ the "dispersion length". It is the maximum
distance the X-type pulse may travel while recovering its initial
longitudinal shape. Obviously, if we want the pulse to reassume
its longitudinal shape at some desired distance $z < L_{\rm disp}$,
we have just to suitably choose the  value of the chirp parameter.

\h The longitudinal shape-evolution described by Eq.(\ref{sol1})
is shown in Fig.1, on adopting the chirp parameter $C=-1$ (and
$\be_2>0$).

\begin{figure}[!h]
\begin{center}
\scalebox{1.1}{\includegraphics{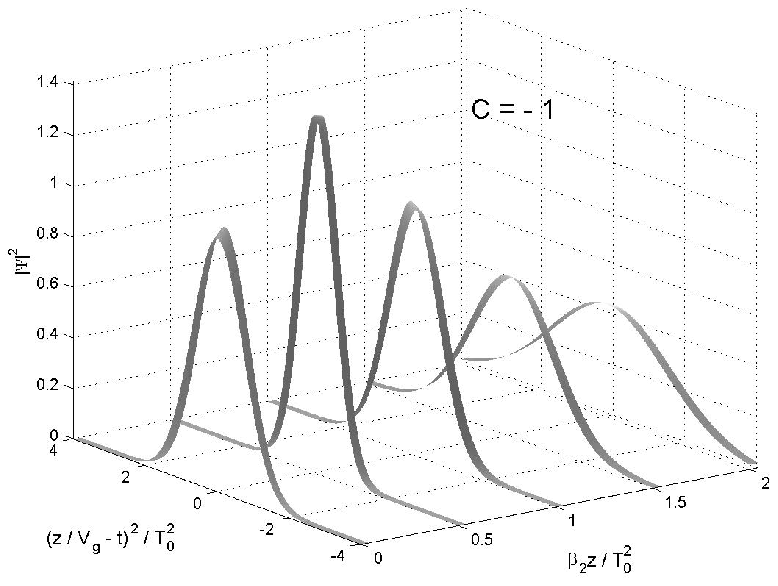}}
\end{center}
\caption{} \label{Fig1}
\end{figure}

\h We can notice that, initially, the pulse suffers a longitudinal
narrowing with an increase of intensity till the position
$z=T_0^2/2\be_2$. After this point, the pulse starts to broaden
decreasing its intensity and recovering its entire longitudinal
shape (width and intensity) at the point $z=T_0^2/\be_2$, as it
was predicted above.

\h Let us emphasize that the property of recovering its own
initial temporal (or longitudinal) width may be verified to exist
also in the case of chirped standard gaussian pulses\footnote{The
chirped gaussian pulse can recover its longitudinal width, but
with a diminished intensity, due to progressive transverse
spreading.}.[14] \ However, the latter will suffer a progressive
transverse spreading, which will not be reversible\footnote{This
problem could be overcome, in principle, by using a lens. But it
would not be a good solution, because it would be necessary a
different lens (besides a different chirp parameter $C$) for each
different value of $Z_{T_1=T_0}$.}. The distance at which a
gaussian pulse doubles its initial transverse width $w_0$ is
$z_{\rm diff} = \sqrt{3}\pi w_0^2/\lambda_0$, where $\lambda_0$ is
the carrier wavelength. Thus, we can see that optical gaussian
pulses with great transverse localization will get spoiled in a
few centimeters or even less.

\h Now we shall show that it is possible to recover also the
transverse shape of  the chirped X-type pulse intensity; actually,
it is possible to recover its entire spatial shape after a distance
$Z_{T_1=T_0}$.

\h To see this, let us go back to our integral solution
(\ref{geral2}), and perform the change of coordinates \
$(z,t) \rightarrow (\Delta z, t_c = z_c/V_g)$, \ with

\

\bb\left\{\begin{array}{l} z \ug z_c + \Delta z\\
\\
 \dis{t=t_c\equiv \frac{z_c}{V_g}} \end{array}\right. \label{zc}\ee

\

where $z_c$ is the center of the pulse ($\Delta z$ is the distance
from such a point), and $t_c$ is the time at which the pulse
center is located at $z_c$. What we are going to do is to compare
our integral solution (\ref{geral2}), when $z_c=0$ (initial pulse),
with that when $z_c=Z_{T_1=T_0}=-2CT_0^2/(\be_2(C_2+1))$.

\h In this way, the solution (\ref{geral2}) can be written, when
$z_c = 0$, as

\

\bb \begin{array}{clcr} \Psi(\rho,z_c=0,\Delta z)
\!\!&=\dis{\frac{T_0\,\,e^{i\be_0\Delta
z}}{\sqrt{2\pi(1+iC)}}\,\int_{-\infi}^{\infi}\,
J_0(k_{\rho}(\om)\rho)\,\exprm\left(\frac{-T_0^2}{2(1+C^2)}(\om-\om_0)^2\right)\times}\\
\\
\
&\;\;\;\dis{\times\,\exprm\left\{i\left[\frac{(\om-\om_0)\Delta
z}{V_g} + \frac{(\om-\om_0)^2\be_2\Delta z}{2} +
\frac{(\om-\om_0)^2T_0^2C}{2(1+C^2)} \right]
\right\}} \end{array}\label{zc0}\ee

\

where we have taken the value $q$ given by Eq.(\ref{s}).

\h To verify that the pulse intensity takes on again its entire
original form at $z_c = Z_{T_1=T_0} = -2\,CT_0^2/ [\be_2(C^2+1)]$,
we can analyze our integral solution at that point,  obtaining:

\

\bb \begin{array}{clcr} \Psi(\rho,z_c=Z_{T_1=T_0},\Delta z)
\!\!&=\dis{\frac{T_0\,e^{i\be_0(z_c-\Delta z'-
\frac{cz_c}{cos\theta\,n(\om_0)V_g})}}{\sqrt{2\pi(1+iC)}}\,\int_{-\infi}^{\infi}\,
J_0(k_{\rho}(\om)\rho)\,\exprm\left(\frac{-T_0^2}{2(1+C^2)}(\om-\om_0)^2\right)\times}\\
\\
\
&\;\;\;\dis{\times\,\exprm\left\{-i\left[\frac{(\om-\om_0)\Delta
z'}{V_g} + \frac{(\om-\om_0)^2\be_2\Delta z'}{2} +
\frac{(\om-\om_0)^2T_0^2C}{2(1+C^2)} \right]
\right\}}\\
\label{zt1}\end{array}\ee

\

where we have put  $\Delta z = -\Delta z'$. \ In this way, one
immediately sees that

\

\bb |\Psi(\rho,z_c=0,\Delta z)|^2 =
|\Psi(\rho,z_c=Z_{T_1=T_0},-\Delta z)|^2 \; . \label{inten}
\ee

\

\h Therefore, from Eq.(\ref{inten}) it is clear that the chirped
optical X-type pulse intensity reassumes its original
three-dimensional form, with just a longitudinal inversion at the
pulse center: The present method being, in this way, a simple and
effective procedure for compensating the effects of diffraction and
dispersion in an unbounded material medium; and a method simpler than
the one of varying the axicon angle with the frequency.

\h Let us stress that we can choose the distance $z =
Z_{T_1=T_0}\leq L_{\rm disp}$ at which the pulse will take on again its
spatial shape by choosing a suitable value of the chirp parameter.

\

\

{\bf 3. -- Analytic description of the transverse pulse behavior
during propagation }\\

%%\h %%Our main goal in this work was already reached, i.e., we have
\h In the previous Section, we have shown that a chirped X-type
pulse can recover its total three-dimensional shape after some
propagation distance in material media, resisting, in
this way, the effects of diffraction and dispersion.

\h We have also obtained an analytic description of the {\em
longitudinal} pulse behavior (for $\rho=0$) during propagation, by
means of Eq.(\ref{sol1}). However, one does not get the same
information about the transverse pulse behavior: We just know that
it is recovered at $z=Z_{T_1=T_0}$).

\h Therefore, it would be interesting if we can get also the {\em
transverse} pulse behavior in the plane of the pulse center $z=V_g
t$.  In that way, we'd get quantitative information about the evolution
of the pulse-shape during its entire propagation.

\h To get this result,  let us go back to Eq.(\ref{geral2}) and
rewrite the transverse wavenumber ($k_{\rho} = (n(\om)\om
\sin\theta)/c$) in the more appropriate form

\

\bb \kr \ug \dis{\sqrt{\frac{n^2(\om)\om^2}{c^2} -
\frac{n^2(\om)\om^2}{c^2}\cos\theta}} \label{kr2} \; . \ee

\

\h To analyze the transverse pulse behavior, it is enough to
expand $n(\om)\om$ at the first order in the vicinity of the carrier
frequency $\om_0$

\bb n(\om)\om \approx A\,u + B \; , \ee

where $u\equiv (\om-\om_0)$ and

\bb A \ug \left(\om_0\frac{\pa n}{\om}|_{\om_0} + n(\om_0)\right)
\ug \frac{c}{V_g\,\cos\theta} \;,\;\;\; B \ug  n(\om_0)\om_0 \; ,
\label{AB} \ee

\h so that $k_{\rho}$ can be written as

\

\bb \kr \approx \frac{\sin\theta}{c}\sqrt{A^2u^2 + 2ABu + B^2}
\label{kr3}
\ee

\

and the Bessel function of the integrand in Eq.(\ref{geral2})
is given by

\bb J_0(\kr(\om)\rho) \ug
J_0\left(\frac{\rho\sin\theta}{c}\sqrt{A^2u^2 + 2ABu +
B^2}\right) \; . \label{j02} \ee

\

\h Now, we use the identity

\bb J_0(mR)\ug J_0(mx)J_0(my) +
2\sum_{p=1}^{\infty}J_p(mx)J_p(my)\cos(p\phi) \; , \ee

\

where $R=\sqrt{x^2+y^2-2xy\cos\phi}$. \ In this way
Eq.(\ref{j02}) can be written

\

\bb
\begin{array}{clcr}
\dis{J_0\left(\frac{\rho\sin\theta}{c}\sqrt{A^2u^2 + 2ABu +
B^2}\right)}\!\!&=
\dis{J_0\left(\frac{\rho\sin\theta}{c}Au\right)J_0\left(\frac{\rho\sin\theta}{c}B\right)+}\\
\\
&\;\;\;\dis{+2\sum_{p=1}^{\infty}J_p\left(\frac{\rho\sin\theta}{c}Au\right)
J_p\left(\frac{\rho\sin\theta}{c}B\right)(-1)^p}
\end{array}\ee

\

\h On using this way of writing the Bessel function, and putting
$z=z_c=V_gt$, we can integrate our solution (\ref{geral2}) and,
after some calculations and recourse to ref.[15], get

\

\bb \begin{array}{l}
\Psi(\rho,z=z_c,t=z_c/V_g)
 \ug \dis{ {\frac {T_0\,\,e^{i\be(\om_0)z}e^{-i\om_0t}}
{\sqrt{2\pi(1+iC)}}} \ \dis{ {\frac {\exprm \left( \dis{
{\frac {-\tan^2\theta\,\rho^2} {8\,V_g^2(-i\be_2
z_c/2 \, + q^2)}}} \right) } {\sqrt{-i\be_2 z_c/2 \, + q^2}}}} }\\
\\
\\
 \times \, \dis{ \left[\Gamma(1/2) J_0 \left( \frac {n(\om_0)\,\om_0\,
\sin\theta\,\rho} {c} \right) I_0 \left( \frac {\tan^2\theta\,\rho^2}
{8\,V_g^2(-i\be_2 z_c/2 \, + q^2)} \right) \right.}\\
\\
\\
 + \left. 
\dis{ 2\sum_{p=1}^{\infty} \frac {2^p\Gamma(p+1/2)\Gamma(p+1)} {\Gamma(2p+1)} \,\,
J_{2p} \left( \frac {n(\om_0)\,\om_0\,\sin\theta\,\rho} {c} \right)
I_{2p} \left( \frac {\tan^2\theta\,\rho^2} {8\,V_g^2(-i\be_2 z_c/2 \, + q^2)}
\right) } \right] 
\\
\ \label{trans1}
\end{array}\ee

\

where we used (\ref{AB}), and $I_p(.)$ is the modified Bessel
function of the first kind of order $p$, quantity $\Gamma(.)$ being the
gamma function and $q$ being given by Eq.(\ref{s}).

\h Equation (\ref{trans1}) describes the transverse pulse behavior
(in the plane $z=z_c=V_gt$) during its whole  propagation. At a first
sight, this solution could appear very complex, but the series
in its right hand side gives a negligible contribution. This fact renders
our solution (\ref{trans1}) of important practical interest.

\h Indeed, from this solution one can see that the transverse pulse width
is governed either by the gaussian function

\bb \exprm\dis{\left(\frac{-\tan^2\theta\,\rho^2}{8\,V_g^2(-i\be_2
z_c/2 \, + q^2)} \right)} \label{gt} \ee

\

or by the Bessel function

\bb J_0\left(\frac{n(\om_0)\,\om_0\,\sin\theta\,\rho}{c}\right) \; ,
\label{bt} \ee

\

whose transverse width are given, respectively, by

\bb \Delta\rho_G (z_c) \ug \frac{2\,c\,\sqrt{(T_0^2 + \be_2 C
z_c)^2 + \be_2^2 z_c^2}}{T_0 \sin \theta \left(n(\om_0)+ \om_0
\frac{\pa n}{\pa \om}|_{\om_0}\right)} \label{drg} \ee

and by

\bb \Delta\rho_B \ug \frac{2.4\,c}{n(\om_0)\om_0\sin\theta} \; ,
\label{drb} \ee

\

where we took advantage of the fact that $V_g=(c/\cos\theta)\pa(\om n(\om))/
\pa\om)|_{\om_0}$, and approximated the first root of the
Bessel function $J_0(.)$ by adopting the value $2.4$.

\h One can see from Eq.(\ref{drg}) that, when $\be_2 C < 0$ (which correspond
to the cases considered in this paper), the gaussian function
(\ref{gt}) will suffer a progressive spreading till $z_c=
-CT_0^2/\be_2(1+C^2)$;  after this point, it will start to spread out with
the distance $z_c$ in an irreversible way, recovering its initial
value $\Delta\rho_G(z_c=0)$ at the point
$z_c=Z_{T_1=T_0}=-2CT_0^2/\be_2(1+C^2)$, as we had forecast in
Section 2.

\h On the other hand, from Eq.(\ref{drb}) we can see that the
transverse width $\Delta\rho_B$ of the Bessel function (\ref{bt})
will be constant during propagation, since the it does not
depend on $z_c$.

\h During its propagation, the pulse's transverse width will be
governed by that one, of the two functions (\ref{gt}) and (\ref{bt}),
{\em which possesses the smaller width.}

\h For example, at the point
$z_c=Z_{T_1=T_0}=-2CT_0^2/\be_2(1+C^2)$, we have that

\bb \frac{\Delta\rho_G(z_c=Z_{T_1=T_0})}{\Delta\rho_B} \ug
\frac{2n(\om_0)}{2.4\left(n(\om_0)+ \om_0 \frac{\pa n}{\pa
\om}|_{\om_0}\right)}\,\,T_0\om_0 \label{dgdb} \ee

\

and we can see that the transverse pulse-width evolution will
depend on the value of $T_0\om_0$, \ the initial temporal
width $T_0$ being related with the pulse frequency bandwidth
$\Delta\om=(1+C^2)^{1/2}/T_0$.

\h Here we can distinguish two different situations:

\

\h {\em When}

$$T_0\om_0 > \frac{2.4\left(n(\om_0)+ \om_0
\frac{\pa n}{\pa \om}|_{\om_0}\right)}{2n\om_0}  $$

\

we have that $\Delta\rho_B < \Delta\rho_G(z_c=Z_{T_1=T_0})$, so that
the transverse pulse width $\Delta\rho$ \ will be governed by the
Bessel function (\ref{bt}), that is, $\Delta\rho=\Delta\rho_B =
2.4\,c /(n(\om_0)\om_0\sin\theta)$ (whose value will remain constant
during the rest of the propagation).\footnote{Obviously, in the case of
finite apertures, one must take into account the finite field depth
of the X pulses. We shall see this in Section 4.}. The pulse
intensity will suffer an intensity decrease due a longitudinal
spreading that starts after $z_c=Z_{T_1=T_0}/2$.

\

\

\h {\em When, by contrast,}

$$T_0\om_0 < \frac{2.4\left(n(\om_0)+ \om_0
\frac{\pa n}{\pa \om}|_{\om_0}\right)}{2n\om_0}  $$

\

we are dealing with ultrashort pulses, while {\em supposing
that $\be_2$ is still enough to describe the material dispersion.}
In these situations, $\Delta\rho_G(z_c=Z_{T_1=T_0}) <
\Delta\rho_B$, so that the transverse pulse width $\Delta\rho$ \ will
be governed by the gaussian function (\ref{gt}), that is,
$\Delta\rho=\Delta\rho_G(z_c)$, given by Eq.(\ref{drg}): And the
pulse will suffer a progressive transverse spread during its
propagation till the distance

$$z_c \ug \frac{\dis{T_0^2|C| + \frac{T_0}{2n(\om_0)\om_0}\sqrt{2.4^2(1+C^2)\left(n(\om_0)+ \om_0
\frac{\pa n}{\pa \om}|_{\om_0}\right)^2 - \,
4\,T_0^2\,n^2(\om_0)\,\om_0^2} }}{|\be_2|(1+C^2)} \; ,
$$

\

where $\Delta\rho_G(z_c)=\Delta\rho_B$. \ From such a point onwards,
the Bessel function (\ref{bt}) starts to rule the pulse
transverse width, $\Delta\rho=\Delta\rho_B=2.4\,c
/(n(\om_0)\om_0\sin\theta)$, which will remain constant i the
rest of the propagation.\footnote{Again, if we consider finite-aperture
generation, one must take into account the finite field
depth of the X pulses, as we shall see in Section 4}. Naturally,
there will be a pulse intensity decrease due to the longitudinal
spreading that takes place after $z_c=Z_{T_1=T_0}/2$.

\h It is interesting to notice that in both cases, the pulse, besides
recovering its full three dimensional shape at $z_c=Z_{T_1=T_0}$,
will reach, after some distance that depends on the considered case, a
constant transverse width $\Delta\rho=2.4\,c
/(n(\om_0)\om_0\sin\theta)$, that is, the same width as that of a Bessel
beam of frequency and axicon angle equal to the carrier frequency and
axicon angle of the chirped X-type pulse.

\h Now, we can use Eq.(\ref{trans1}) to show the transverse
behavior of a X-Type pulse propagating, e.g., in fused Silica
($SiO_2$) with the following characteristics: $T_0=0.4 \;$ps,
$\la_0=0.2\mu \;$m, and axicon angle $\theta=0.002\;$rad. For
evaluating\footnote{Here, it is not necessary to evaluate $\be_2$,
because we can use the normalized quantity $\be_2z\,/T_0^2$.} the
quantity $\be_1$, we need the refractive index function $n(\om)$.
Far from the medium resonances, it can be approximated by the
well-known Sellmeier equation[14]

\

\bb n^2(\om) \ug
\dis{1\,+\,\sum_{j=1}^{N}\,\frac{B_j\,\om_j^2}{\om_j^2- \om^2}} \;
, \label{Selm}\ee

\

where $\om_j$ are the resonance frequencies, $B_j$ the strength of
the $j$th resonance, and $N$ the total number of the material
resonances that appear in the frequency range of interest. For our
purposes it is appropriate to choose $N=3$, which yields, for bulk
fused Silica, the values[14] $B_1=0.6961663$; \ $B_2=0.4079426$; \
$B_3=0.8974794$; \ $\lambda_1=0.0684043 \; \mu \;$m; \
$\lambda_2=0.1162414 \; \mu\;$m; \ and $\lambda_3=9.896161 \;
\mu\;$m.
%%With this we can obtain that $\be_1=XX$ and $\be_2=XX$.

\h Figure (2) shows the evolution of the transverse shape of such a
pulse with chirp parameter $C=-1$ \  (so that $Z_{T_1=T0}=L_{\rm
Disp}=T_0^2/\be_2$). One can notice that the pulse maintains its
transverse width $\Delta\rho=2.4\,c/(n(\om_0)\om_0
\sin\theta)=24.635\mu \;$m during its entire propagation due the fact
that $T_0\om_0>>1$; however, the same does not occur with the pulse
intensity. Initially the pulse suffers an increase of intensity
till the position $z_c=T_0^2/2\be_2$; after this point, the
intensity starts to decrease, and the pulse recovers its entire
transverse shape at the point $z_c=T_0^2/\be_2$, as it
was expected by us.  Here we have skipped the series on the right hand
side of Eq.(\ref{trans1}), because, as we already said, it yields a
negligible contribution.

\

\begin{figure}[!h]
\begin{center}
\scalebox{1.1}{\includegraphics{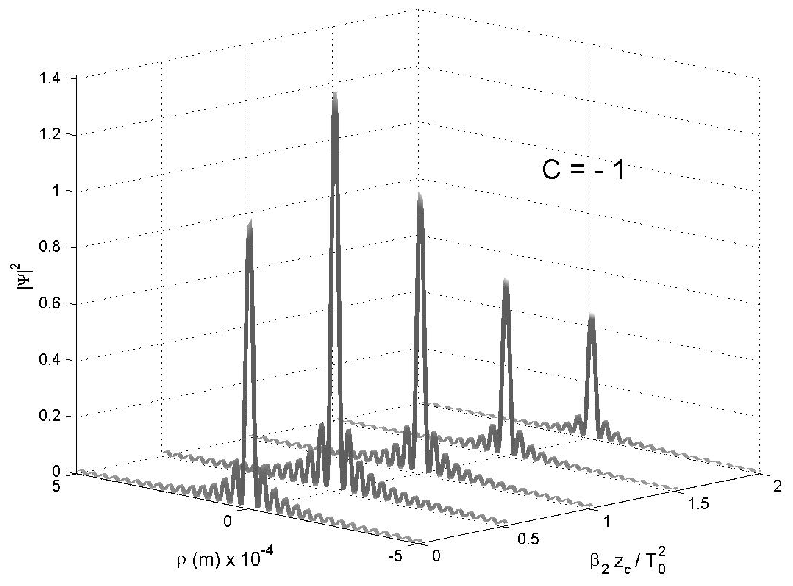}}
\end{center}
\caption{} \label{Fig2}
\end{figure}

\

\newpage

{\bf 4. -- Advantages and limitations of using chirped X-type waves}\\

\h The main advantage of the present method is its simplicity. A
chirped optical X-type pulse with a constant axicon angle can be
generated in a very simple way, by using an annular slit localized
at the focus of a convergent lens, and illuminating that slit with
a chirped optical gaussian pulse (where the chirp can be
controlled directly in the laser modulation).

\h However, we should recall that we have taken into
account the dispersion effects till their second order. When the
pulse wavelength nearly coincides with the zero dispersion
wavelength (i.e., $\be_2 \approx 0$), or when the pulse (depending on
the material) is ultrashort, then it is necessary to include the
third order dispersion term $\be_3$, which in those cases will
provide the dominant GVD effect.  Consequently, the use of
chirped optical X-type pulses might not furnish in those case the same
results as shown above.  A good option in such cases would be varying the
axicon angle with the frequency.

%%In the cases of ultrashort pulses, or when the pulse wavelength
%%nearly coincides with the zero dispersion wavelength (ie, $\be_2
%%\approx 0$), it is necessary to include the third order dispersion
%%term $\be_3$, which in those cases will provide the dominant GVD
%%effect. Correspondingly, the use of chirped optical X-type pulses
%%might not furnish the same results shown above.  A good option in
%%such cases would be varying the axicon angle with the frequency.

\h Let us also recall that a Bessel beam, generated by finite
apertures (as it must be, in any real situations), maintains its
nondiffracting properties till a certain distance only (called its
{\em field depth\/}), given by

\bb L_{\rm diff} \ug  \frac{R}{\tan\theta} \; , \label{ldif}\ee

where $R$ is the aperture radius and $\theta$ is the axicon angle.
We call this distance $L_{\rm diff}$ to emphasize that it is the
distance along which a X-type wave can resist the diffraction
effects.

\h So, since our chirped X-type pulse is generated by a frequency
superposition of BBs with the same axicon angle, our pulse will be able
to take on again its shape at the position $z = Z_{T_1=T_0} \leq L_{\rm disp}$,
if those BBs themselves are able to reach such a point resisting to the
diffraction effects; in other words, if

\

\bb Z_{T_1=T_0}\leq L_{\rm diff} \;\;\; \rightarrow \;\;\;
\frac{-2CT_0^2}{\be_2(C^2+1)}\leq \frac{R}{\tan\theta} \ee

\

\h This fact leads us to conclude that the dispersion length
$L_{\rm disp}$ and the diffraction length $L_{\rm diff}$  play
important roles in the applications of chirped X-type pulses. Such
roles are exploited (and summarized) in the following two cases:

\

\h {\em First case}:

\h When  $L_{\rm disp} \leq  L_{\rm diff}$:

\h In this case the {\em dispersion} plays the critical role in the
pulse propagation, and we can ensure pulse fidelity till a maximum
distance given by $z = L_{\rm disp} = T^2_0/\be_2$. More
specifically, we can choose a distance $z = Z_{T_1=T_0} \leq
L_{\rm disp}$ at which the pulse will reassume its whole spatial
shape by choosing the correct value of the chirp parameter.

\

\h {\em Second case}:

\h When  $L_{\rm diff} \leq L_{\rm disp}$ :

\h In this case the {\em diffraction} plays the critical role in the
pulse propagation. When this occurs, we can emit in the dispersive
medium a chirped X-type pulse that takes on again its entire spatial
shape after propagating till the maximum distance, given by $z =
L_{\rm diff}$. To do this, we have to choose the correct value of
the chirp parameter.

\h Let us suppose that we want the pulse intensity to reach the
maximum distance $L_{\rm diff}$  with the same spatial shape as in
the beginning.   We should have, then:

\

\bb \frac{-2CT_0^2}{\be_2(C^2+1)} \ug \frac{R}{\tan\theta} \ .
\label{condC} \ee

\

\h Once  $T_0$, $\be_2$, \ $R$ and \ $\theta$  are known, we can use
Eq.(\ref{condC}) to find the correct value of the chirp parameter.

\

\

{\bf 5. -- Comparison with the ordinary chirped gaussian pulses}\\

\h It would be interesting if some comparison was made among the
chirped X-type pulses and the ordinary chirped gaussian pulses.

\h {\em In this comparison we will deal, with respect to the X-pulses,
with the situation given by the first case in Section $4$; we will use
the condition $L_{\rm diff} \geq 2\,L_{\rm disp}$. \ Then,
we can approximate a chirped X-type pulse, generated by
finite apertures (finite energy), by having recourse to the equations previously
obtained, at least till the distance $2L_{\rm disp}$.} \ On
the contrary, one ought to use numerical simulations for obtaining the
pulse behavior till that distance.

%%This situation is very acceptable because, in general, for pulses
%%with narrow transverse width generated by finite apertures we have
%%that $(L_{\rm diff})_{X}>>(L_{\rm diff})_{\rm Gauss}$

\h Now, on using the paraxial approximation, a chirped gaussian pulse
of initial transverse width $w_0$, initial temporal width $T_0$,
chirp parameter $C$ and carrier frequency $\om_0$, and
propagating in a material medium with refractive index $n(\om)$,
can be written as

\

\bb \Psi_{\rm gaussian}(\rho,z,t) \ug \frac{\dis{{\rm
exp}\left(\dis{\frac{-\rho^2}{w_0^2 + 2\,i\,z/k_0}}\right)\,{\rm
exp}\left(\dis{\frac{(z-V_g\,t)^2(1+i\,C)}{2\,V_g^2\,T_0^2(1-i\,(1+i\,C)\be_2\,z/T_0^2)}}\right)
}}{\dis{\left(1+\frac{2\,i\,z}{w_0^2\,k_0}\right)\,\sqrt{1-i\,(1+i\,C)\,\frac{\be_2\,z}
{T_0^2}}}} \; , \label{gauss}\ee

\

where $k_0=2\pi/\lambda_0=\om_0/c$, \ $V_g=1/\be'(\om)|_{\om_0}$,\
\ $\be_2=\be''(\om)|_{\om_0}$ \ with \ $\be(\om)=n(\om)\om/c$.

\h To perform the comparison, we will consider the following
situation: Both pulses possess the same spot-size,\footnote{We call
spot-size the transverse width, which for gaussian pulses is
the transverse distance from the pulse center to the position at which
the intensity falls down by a factor $1/e$; and, in the case of the
considered X-pulses,
the transverse distance from the pulse center to where the first zero
of the intensity occurs.}  the same carrier frequency and temporal
width, the same chirp parameter, and propagate inside fused silica.
The values of the group velocity, and the group velocity dispersion, of each
pulse will be calculated by using the previous
relations,\footnote{Remembering that $(V_g)_X=(V_g)_{\rm
Gauss}/\cos\theta$ and $(\be_2)_X=\cos\theta (\be_2)_{\rm Gauss}$.}
the refractive index $n(\om)$ being approximated by the
Sellmeier equation (\ref{Selm}).

\h Thus, we may consider a gaussian pulse with: $\lambda_0=0.2\mu \;$m, \
$T_0=0.4 \;$ps, \ $C=-1$ and an initial transverse width (spot-size)
$w_0=0.117 \;$mm; Figure 3 shows the longitudinal and
transverse behavior of this pulse during propagation.

\h We can see from Fig.3a that the chirped gaussian pulse may
recover its longitudinal width, but with an intensity decrease, at
the position given by $z = Z_{T_1=T_0}=L_{\rm disp} = T^2_0/\be_2$
(because $C=-1$), which, in this case, is equal to $0.373 \;$m. Its
transverse width, on the other hand, suffers a progressive
spreading and, in the considered case, {\em doubles} its value\footnote{One
can easily verify
that in this case the distance $L_{\rm disp}$ coincides with
$z_{\rm diff}=\sqrt{3}\pi w_0^2/\lambda_0$, which is the distance
where a {\em gaussian pulse} doubles its transverse width.}
at the position $z = L_{\rm disp} = T^2_0/\be_2=0.373 \;$m, as
one can see from Fig.3b.

\

\begin{figure}[!h]
\begin{center}
\scalebox{1.8}{\includegraphics{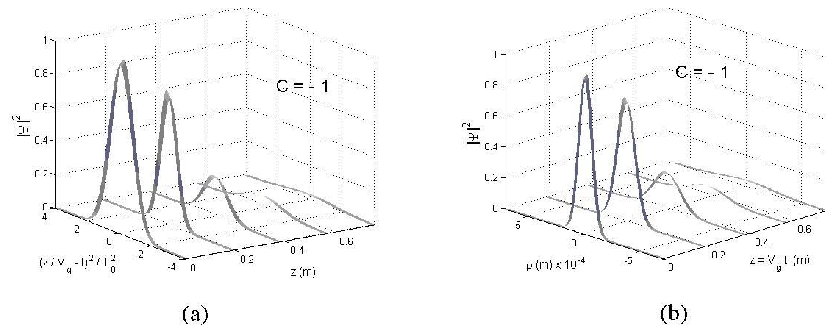}}
\end{center}
\caption{} \label{Fig3}
\end{figure}

\

\h Considering now in the same medium (fused silica) a chirped
X-type pulse, with the same $\lambda_0$, $T_0$, $C$ and with an
axicon angle $\theta=0.00084 \;$rad, which correspond to an initial central
spot with $\Delta\rho_0=0.117 \;$mm (the same as for gaussian pulses), we
get, during propagation, the longitudinal and transverse shapes
represented by Fig.4.

\

\begin{figure}[!h]
\begin{center}
\scalebox{1.8}{\includegraphics{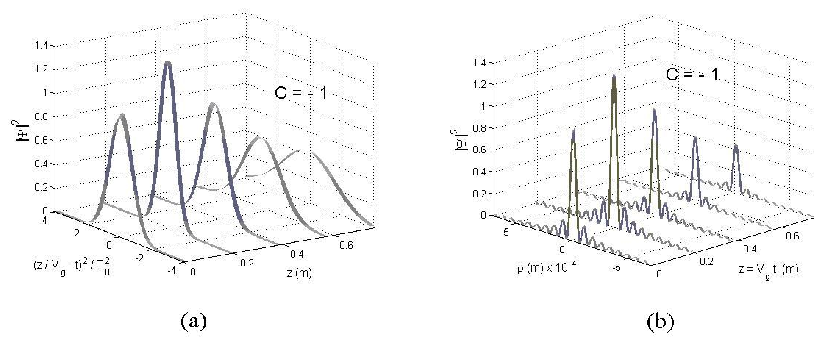}}
\end{center}
\caption{} \label{Fig4}
\end{figure}

\

\h As one can see from Figs.4a and 4b, the pulse recovers totally
its longitudinal and transverse shape at the position $z = L_{\rm
Disp} = T^2_0/\be_2=0.373 \;$m.

\h Again, we can notice that, during its entire propagation, the
chirped X-type pulse maintains its transverse width
$\Delta\rho=2.4\,c/(n(\om_0)\om_0 \sin\theta)=0.117 \;$mm \ (because
$T_0\om_0>>1$),  decreasing however its intensity after $z =
T_0^2/2\be_2$ because of the progressive longitudinal broadening that
occurs after that point. In the case of a finite-aperture
generation, this behavior will be approximately maintained till
$z=L_{\rm diff}=R/\tan\theta$.

\h Considering an aperture of radius $R=1 \;$mm, the chirped X-type
pulse considered above would have $L_{\rm diff}=1.186 \;$m.

\

\

{\bf 6. -- Conclusions}\\

\h In this paper we have proposed the use of chirped optical
X-shaped pulses in dispersive media to overcome the problems of both
diffraction and dispersion. We have shown that the dispersion and the
diffraction length, $L_{\rm disp}$ and $L_{\rm diff}$,
respectively, play essential roles on the recovering  of the
pulse-intensity shape.

\

{\bf Acknowledgements}\\

\h The authors are very grateful to Amr Shaarawi for continuous
discussions and collaboration. Useful discussions are moreover
acknowledged with with J.M.Madureira,
S.Zamboni-Rached, V.Abate, M.Brambilla,
C.Cocca, R.Collina, C.Conti, G.C.Costa, G.Degli Antoni,
G.Kurizki, G.Marchesini, D.Mugnai, M.Pernici, V.Petrillo, A.Ranfagni,
G.Salesi, J.W.Swart, M.T.Vasconselos and M.Villa.\\

\

{\bf 7. -- Figure Captions}\\

{\bf Fig.1 --} Longitudinal-shape evolution of a chirped
X-shaped pulse with $C=-1$. One can see that such a pulse recovers its
full longitudinal shape at the position $z = L_{\rm disp} =
T^2_0/\be_2$.

\

{\bf Fig.2 --} Transverse-shape evolution of a chirped
X-type pulse with $C=-1$, \ $T_0=0.4 \;$ps, \ $\la_0=0.2\mu \;$m, and
axicon angle $\theta=0.002 \;$rad, which correspond to an initial transverse
width of $\Delta\rho=24.63\mu \;$m. \ One can see that the pulse
recovers its entire transverse shape at the distance $z = L_{\rm
Disp} = T^2_0/\be_2$: Which, in this case, is equal to
$0.373 \;$m.

\

{\bf Fig.3 --} \ (a): Longitudinal-shape evolution of a chirped
gaussian pulse propagating in fused silica, with
$\lambda_0=0.2\mu \;$m, \ $T_0=0.4 \;$ps, \ $C=-1$ and initial transverse
width (spot-size) $\Delta\rho_0=0.117 \;$mm. \ \ (b): Transverse-shape
evolution for the same pulse.

\

{\bf Figs.4.} \ (a): Longitudinal-shape evolution of a chirped
X-type pulse, propagating in fused silica with
$\lambda_0=0.2\mu \;$m, \ $T_0=0.4 \;$ps, \ $C=-1$ and axicon angle
$\theta=0.00084 \;$rad, which correspond to an initial transverse width of
$\Delta\rho_0=0.117 \;$mm. \ \ (b): Transverse-shape evolution for the same
pulse.

\

\

{\bf References:}\hfill\break

[1] I.M.Besieris, A.M.Shaarawi and R.W.Ziolkowski, ``A
bi-directional traveling plane wave representation of exact
solutions of the scalar wave equation", J. Math. Phys.,
\textbf{30}, 1254-1269 (1989).

\vspace{0.4cm}

[2] J.-y.Lu and J.F.Greenleaf, ``Nondiffracting X-waves: Exact
solutions to free-space scalar wave equation and their finite
aperture realizations", IEEE Trans. Ultrason. Ferroelectr. Freq.
Control, \textbf{39}, 19-31, (1992).

\vspace{0.4cm}

[3] For a review, see: E.Recami, M.Zamboni-Rached, K.Z.N\'obrega,
C.A.Dartora, and H.E.Hern\'andez-Figueroa, ``On the localized
superluminal solutions to the Maxwell equations", IEEE Journal of
Selected Topics in Quantum Electronics, \textbf{9}, 59-73 (2003);
and references therein.

\vspace{0.4cm}

[4] M.Zamboni-Rached, E.Recami, and H.E.Hern\'andez-Figueroa,
``New localized Superluminal solutions to the wave equations with
finite total energies and arbitrary frequencies", European
Physical Journal D, \textbf{21}, 217-228 (2002).

\vspace{0.4cm}

[5] See, e.g., M.Zamboni-Rached, E.Recami, and F.Fontana,
``Localized Superluminal solutions to Maxwell equations
propagating along a normal-sized waveguide", Physical Review E,
\textbf{64}, article no.066603 (2001).

\vspace{0.4cm}

[6] H.S\~onajalg, P.Saari, ``Suppression of temporal spread of
ultrashort pulses in dispersive media by Bessel beam generators"
Optics Letters, \textbf{21}, 1162-1164 (1996).

\vspace{0.4cm}

[7] H.S\~onajalg, M.Ratsep, P.Saari, ``Demonstration of the
Bessel-X pulse propagating with strong lateral and longitudinal
localization in a dispersive medium " Optics Letters, \textbf{22},
310-312 (1997).

\vspace{0.4cm}

[8] M. Zamboni-Rached, K.Z. N\'obrega, H.E.Hern\'andez-Figueroa,
and E. Recami,  ``Localized Superluminal solutions to the wave
equation in (vacuum or) dispersive media, for arbitrary
frequencies and with adjustable bandwidth", Optics Communications,
\textbf{226}, 15-23 (2003).

\vspace{0.4cm}

[9] C.Conti, and S.Trillo, ``Paraxial envelope X waves", Optics
Letters, \textbf{28}, 1090-1093 (2003).

\vspace{0.4cm}

[10] M.A.Porras, G. Valiulis and P. Di Trapani, ``Unified
description of Bessel X waves with cone dispersion and tilted
pulses", Physical Review E, \textbf{68}, article no.016613 (2003).

\vspace{0.4cm}

[11] M.A.Porras, and I.Gonzalo, ``Control of temporal
characteristics of Bessel-X pulses in dispersive media", Optics
Communications, \textbf{217}, 257-264 (2003).

\vspace{0.4cm}

[12] M.A.Porras, R.Borghi, and M.Santarsiero, ``Suppression of
dispersion broadening of light pulses with Bessel-Gauss beams",
Optics Communications, \textbf{206}, 235-241 (2003).

\vspace{0.4cm}

[13] S. Longhi, ``Spatial-temporal Gauss-Laguerre waves in
dispersive media", Physical Review E, \textbf{68}, article
no.066612 (2003).

\vspace{0.4cm}

[14] G.P.Agrawal, {\em Nonlinear Fiber Optics} (Academic Press;
San Diego, CA, 1995).

\vspace{0.4cm}

[15] I.S.Gradshteyn, and I.M.Ryzhik, {\em Integrals, Series and
Products}, 4th edition (Acad. Press; New York, 1965).

\end{document}